\documentclass{article}
\usepackage{spconf,amsmath,graphicx,hyperref}

\usepackage{comment}
\usepackage{makecell}
\usepackage{cite}
\usepackage{enumitem}
\usepackage{array}  
\usepackage{pgfplots}
\pgfplotsset{compat=1.18}
\usepackage{multirow}
\usepackage{amsmath,amssymb,amsfonts}
\usepackage{algorithmic}
\usepackage{graphicx}
\usepackage{textcomp}
\usepackage{xcolor}

\usepackage{setspace}
\usepackage{etoolbox}

\vspace{-5mm}
\title{Towards Noise-Robust Speech Inversion through Multi-Task Learning with Speech Enhancement}

\name{Saba Tabatabaee$^{1}$\thanks{This work was supported by NSF Grant No. BCS2141413.},
      Carol Espy-Wilson$^{1}$}
\address{
    $^{1}$Department of Electrical and Computer Engineering, University of Maryland, College Park, USA \\
}
\begin{document}

\makeatletter
%\begin{comment}
    
\twocolumn[%
\begin{@twocolumnfalse}%
\vspace{-0.5em}
\begin{center}
\small
\textcopyright~ 20XX IEEE. Personal use of this material is permitted. Permission from IEEE must be obtained for all other uses, in any current or future media, including reprinting/republishing this material for advertising or promotional purposes, creating new collective works, for resale or redistribution to servers or lists, or reuse of any copyrighted component of this work in other works.
\end{center}

\vspace{1em}%
\end{@twocolumnfalse}%
]
\makeatother
%\end{comment}
\maketitle

\begin{abstract}
Recent studies demonstrate the effectiveness of Self Supervised Learning (SSL) speech representations for Speech Inversion (SI). However, applying SI in real-world scenarios remains challenging due to the pervasive presence of background noise. We propose a unified framework that integrates Speech Enhancement (SE) and SI models through shared SSL-based speech representations. In this framework, the SSL model is trained not only to support the SE module in suppressing noise but also to produce representations that are more informative for the SI task, allowing both modules to benefit from joint training. At a Signal-to-Noise Ratio of $–$5 dB, our method for the SI task achieves relative improvements over the baseline of 80.95\% under babble noise and 38.98\% under non-babble noise, as measured by the average Pearson product–moment correlation across all estimated parameters. 
\end{abstract}

\begin{keywords}Speech inversion, speech enhancement, multi-task learning, noisy speech.
\end{keywords}

\vspace{-2mm}
\section{Introduction}
\vspace{-2mm}
Speech production involves precise coordination of the lips, tongue, jaw, velum, and glottis to shape the acoustic signal \cite{stevens2000acoustic}. Speech Inversion (SI) aims to recover this hidden articulatory activity by mapping speech onto vocal Tract Variables (TVs), which represent functional synergies among articulators rather than discrete positions \cite{browman1986towards}. The oral TVs are characterized by the degree and location of constrictions formed by the lips, tongue body, and tongue tip (see Figure 1).%The SI system reconstructs these as continuous time-varying trajectories.
%saltzman1989dynamical

\begin{figure}[htbp]
    \hfill
     \includegraphics[width=0.49\textwidth, height=0.15\textheight]{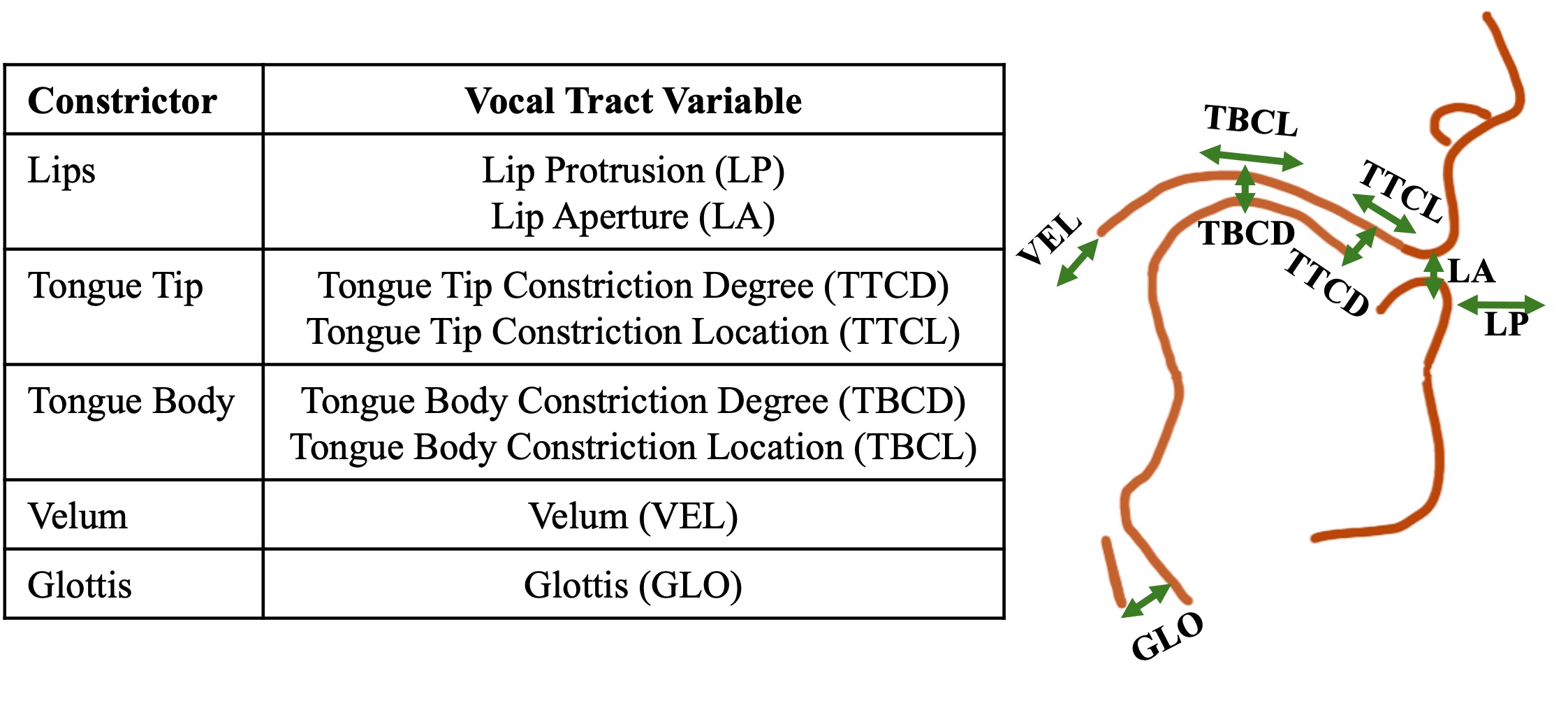}  % Adjust width as needed
     \vspace{-9mm} 
    \caption{A visual illustration of the vocal tract variables.}
    \label{fig:si}
    \vspace{-2mm} 
\end{figure}

In addition to oral sounds, speech also includes nasal sounds, which are produced by partial or complete opening of the Velopharyngeal Port (VP). The degree of VP constriction regulates nasality by controlling airflow and acoustic coupling between the nasal and oral cavities through coordinated movements of the velum and pharyngeal walls. Direct measurement of VP activity is invasive, costly, and requires trained specialists. As a non-invasive alternative, nasalance, defined as the ratio of acoustic energy emitted through the nasal and oral cavities, serves as an indirect measure of VP constriction and has been validated through high-speed nasopharyngoscopy imaging \cite{siriwardena2024speaker}. In this study, nasalance is referred to as the VEL TV, as it serves as an acoustic proxy for velum movement patterns.

%Simultaneous estimation of oral TVs and VEL TV has been shown to improve the accuracy of both in SI systems \cite{tabatabaee25b_interspeech}. Furthermore, integrating Source Features (SFs) extracted from the speech signal, including aperiodicity (non-periodic energy), periodicity (periodic energy), and fundamental frequency (F0), as proxies for glottal control has been shown to improve the estimation of both oral TVs \cite{tabatabaee25b_interspeech} and VEL TV \cite{tabatabaee2025acoustic}. Therefore, in this study, we adopt the recently developed SI system from \cite{tabatabaee25b_interspeech}, which simultaneously estimates oral TVs, VEL TV, and three SFs.
Our investigation, supported by ablation studies \cite{tabatabaee25b_interspeech,tabatabaee2025acoustic}, shows that jointly estimating oral TVs, the VEL TV together with source features (SFs), including fundamental frequency (F0), periodicity (periodic energy), and aperiodicity (aperiodic energy) leads to improvements in their estimation. This is because these variables are inherently related and provide complementary information. Therefore, in this study, we adopt the recently developed SI system from \cite{tabatabaee25b_interspeech}, which simultaneously estimates oral TVs, VEL TV, and three SFs.

% noise 
SI systems have a wide range of applications, including assessment for children with speech disorders \cite{benway25_interspeech,benway2025perceptual,tabatabaee2025acoustic}. However, deploying SI systems in real-world environments is challenging, as background noise can degrade performance if the system has not been adapted to such conditions. A previous study \cite{siriwardena2023audio} demonstrated that adapting a SI system to noisy conditions can be achieved by fine-tuning with noise-contaminated speech signals. However, employing a Speech Enhancement (SE) model as a pre-processing step has been shown to improve robustness across various speech processing tasks \cite{tzeng2025noise, shi2024waveform, shahrebabaki2021acoustic} and to outperform approaches that rely solely on fine-tuning the target system with contaminated speech \cite{shi2024waveform, shahrebabaki2021acoustic}. % In particular, prior work on SI systems highlights that incorporating SE as a pre-processing step enables more effective adaptation to noise than conventional fine-tuning with noisy speech \cite{shahrebabaki2021acoustic}. 
Therefore, in this study, we leverage a SE model to improve the adaptation of an SI system to noisy conditions under two different scenarios: (1) fine-tuning the SI system on audio pre-processed by the SE model, and (2) jointly training the SI and SE models within a multi-task learning framework. To the best of our knowledge, this study is the first to investigate multi-task learning of SI and SE models to improve SI robustness under noisy conditions.

%Previous studies on SI systems have shown that fine-tuning with contaminated audio \cite{siriwardena2023audio}, as well as using a SE model as a pre-processing step \cite{shahrebabaki2021acoustic}, are both effective strategies for adapting SI systems to noisy conditions. In this study, we investigate two strategies for adapting SI system to environmental noise using SE model: (1) fine-tuning the SI system on enhanced audio produced by a SE model, and (2) jointly training SI and SE in a multi-task learning framework. To the best of our knowledge, this is the first study to explore multi-task learning for jointly modeling SI and SE.

% ssl 
Recent advances in Self Supervised Learning (SSL) speech models have significantly improved feature extraction for SI systems, surpassing traditional acoustic representations such as MFCCs \cite{cho2023evidence}. Among these models, WavLM-Large has been found to outperform other SSL models, including HuBERT-Large, on SI tasks \cite{cho2023evidence, tabatabaee25b_interspeech}. Furthermore, several studies have explored the application of SSL models to SE tasks, reporting promising results \cite{huang2022investigating, khan2024exploiting}. Motivated by these findings, we employ the WavLM-Large pretrained SSL model \cite{chen2022wavlm} for the SI and SE systems to extract richer, more informative representations from the speech signal.

\textbf{Our key contributions} in the current work are as follows:
\begin{itemize}[itemsep=0pt, topsep=0pt, parsep=0pt, partopsep=0pt]
    \item Introduction of a novel multi-task learning framework jointly modeling SE and SI to estimate enhanced audio along with oral TVs, VEL TV, and three SFs.
    \item Comparative analysis of employing a SE model as a pre-processing step for a SI model versus jointly integrating SE and SI models within a multi-task learning framework.
    \item Evaluation of SI system robustness under babble and non-babble background noise across varying Signal-to-Noise Ratio (SNR) levels.
\end{itemize}
\vspace{-2mm}
\section{Data prepration}
\vspace{-2mm}
\subsection{XRMB dataset}
\vspace{-2mm}
The University of Wisconsin XRMB dataset \cite{westbury1994speech} comprises speech recordings accompanied by point-source trajectories of the lips, tongue tip, and tongue body, captured using a rasterized X-ray microbeam tracking system. %Pellets are placed on articulators, including the upper lip (UL), lower lip (LL), ventral tongue (T1), mid-ventral tongue (T2), mid-dorsal tongue (T3), dorsal tongue (T4), mandibular incisor (MANi), and mandibular molar (MANm). 
%After removing tracking errors, the dataset includes recordings from 46 native English speakers (21 males and 25 females), totaling approximately 4 hours of speech. By applying the reconstruction methods described in \cite{attia2023masked}, much of the corrupted data is recovered, extending the total duration to approximately 5.74 hours. 
The dataset includes recordings from 46 native English speakers (21 males and 25 females), totaling approximately 5.74 hours of speech. The dataset is partitioned into training, development, and test sets in a speaker-independent manner, as detailed in Table 1. 
The original X–Y coordinates of the pellets were mapped to the TVs using a geometric transformation, as described in \cite{attia2024improving}. Because the TVs are based on relative measures, this mapping normalizes variability arising from anatomical differences among speakers. The processed XRMB dataset, following geometric transformation, comprises six oral TVs: LA, LP, TBCL, TBCD, TTCL, and TTCD (details are provided in Figure 1). The VEL TV was extracted from the XRMB speech signals using the method proposed in \cite{tabatabaee25b_interspeech}, while ground-truth values for the three SFs were obtained from speech signals using the APP detector \cite{deshmukh2005use}.% The oral TVs, the VEL TV, and the three SFs were sampled at 100 Hz.
\vspace{-4mm} 
\begin{table}[ht!]
\caption{Description of the XRMB dataset using speaker-independent splits.}
\footnotesize % Applies to the entire table
\centering % Centers the table within the column span
\label{speaker_verification_models}
\resizebox{\columnwidth}{!}{
    \begin{tabular}{|l|c|c|c|}
\hline
Split & Number of subjects & Number of Utterance & Time (hours) \\
\hline
 Train &36& 5801& 4.47  \\
 \hline
  Development & 5& 730&  0.63 \\
  \hline
   Test &5&754&  0.64 \\
 \hline
    \end{tabular}
       }
       \vspace{-4mm}
\end{table}
\vspace{-2mm} 
\subsection{Data augmentation for XRMB dataset}
\vspace{-2mm} 
For data augmentation, two categories of environmental noise were employed: (1) babble noise and (2) non-babble background noise. The training and development sets were augmented using the same noise data, whereas the test set was augmented with a separate set of noise data to assess SI and SE models performance under unseen noise conditions. \\
\textbf {Noise augmentation of train and development sets}:
For non-babble background noise, we used the MUSAN dataset \cite{Snyder2015MUSANAM}, and babble noise was generated from the VoxCeleb1 test set (Vox1-test) \cite{nagrani2020voxceleb} by overlapping speech segments from 5–20 speakers. The XRMB training and development sets were augmented with three noise conditions: babble, non-babble background noise, and a combination of both, then merged into a single dataset. For each condition, clean XRMB audio was contaminated with noise at randomly selected SNRs between 0 and 10 dB.\\
\textbf{Noise augmentation of test set:}
To evaluate the effect of background noise on SI and SE performance, the XRMB test set was contaminated with two noise types: (1) babble noise, simulated by overlapping speech from 5–20 speakers in the Librispeech test-clean dataset \cite{panayotov2015librispeech}, and (2) non-babble background noise from the DNS-Challenge dataset \cite{reddy2021interspeech}. Each noise type was added at four SNR levels: $-$5 dB, 0 dB, 5 dB, and 10 dB.
\begin{comment}
\subsection{VCTK-DEMAND Dataset}
The SE model was developed using the widely used VCTK-DEMAND premixed dataset \cite{botinhao2016investigating}. The development set contained 11,572 utterances and 9.39 hours of speech from 28 speakers (14 male and 14 female) with four signal-to-noise ratios (SNRs) of 15, 10, 5, and 0 dB. The test set consisted of 824 utterances (0.58 hours) from two speakers (a male and a female) at SNRs of 17.5, 12.5, 7.5, and 2.5 dB. The development data was randomly partitioned into training and validation subsets by a 95\% to 5\% ratio. The training set contained 10,993 utterances (8.92 hours) and the validation set included 579 utterances (0.47 hours). 
\end{comment}
\vspace{-6mm}
\section{Methodology}
\vspace{-2mm}

In this study, both the SI and SE systems use WavLM-Large, a SSL model pre-trained on noisy speech, which provides more robust representations than other SSL models. Models are optimized with AdamW and early stopping (patience of five epochs) to prevent overfitting. To stabilize training and protect SSL representations from randomly initialized task heads, a Two-Stage Training (TST) strategy is employed. In the first stage, the task-specific models (SE, SI, or both, depending on the setup) are trained with WavLM-Large frozen at a learning rate of $5 \times 10^{-4}$. After convergence, all components, including WavLM-Large, are fine-tuned jointly with a reduced learning rate of $5 \times 10^{-5}$. All models (SE-Base, SISE-P, and the proposed SISE-M) are trained using this TST strategy.
\vspace{-1mm}
\begin{figure}[htbp]
    \hfill
    \vspace{-3mm}
    
     \includegraphics[width=0.49\textwidth, height=0.25\textheight]{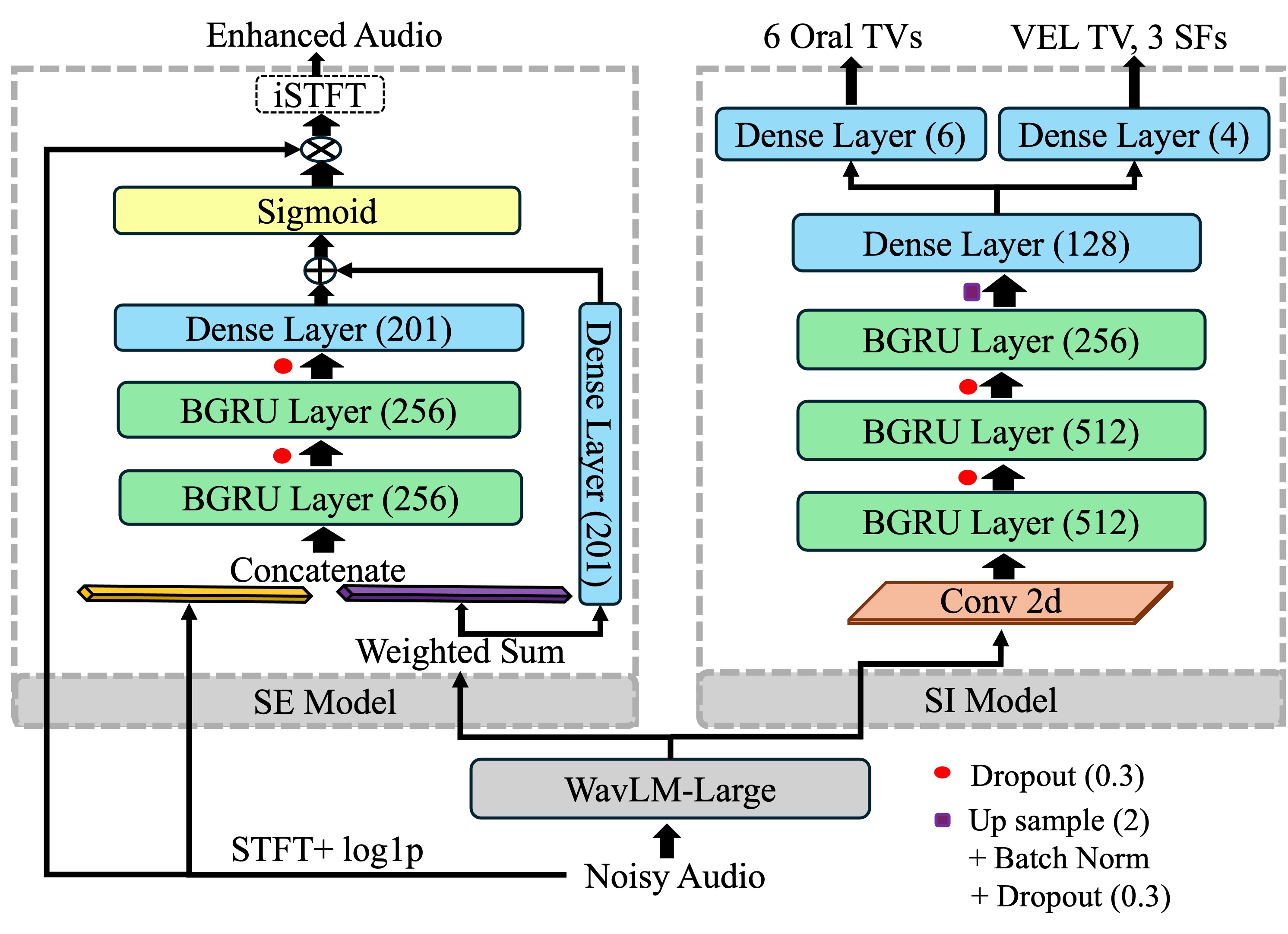}  % Adjust width as needed
     \vspace{-6mm}
    \caption{Architecture of the proposed SISE-M model.}
    
    \label{fig:si}
    \vspace{-6mm} 
\end{figure}
\subsection{SE model}
\vspace{-2mm}
As shown in Figure 2, a weighted sum of features from all 25 WavLM-Large layers is computed to capture contextual information from noisy audio. This sum is concatenated with the spectral representation (log1p-compressed STFT magnitude) and fed through two 256-unit bidirectional GRU layers, followed by a 201-unit dense layer. The output of the dense layer is added to that of a 201-unit dense layer applied to the weighted WavLM-Large features. A sigmoid activation is applied to generate a mask, which is then multiplied with the spectral representation to suppress noise. The masked magnitude is decompressed and converted back to the time domain using the inverse STFT to produce the enhanced audio. 

%For training the SE model, noisy data are pre-processed using Perceptual Contrast Stretching (PCS) \cite{chao2022perceptual}, a spectral technique that adjusts the magnitude spectrum according to perceptual importance of different frequency bands in the human auditory system. PCS is applied to the noisy training, development, and test data, while the clean test data remain unprocessed to ensure fair evaluation. 
The SE model is trained with the combined losses of Weighted Signal-to-Distortion Ratio (WSDR), Compressed Magnitude Spectrum (CMS), and Multi-Resolution STFT (MRS), as shown in equation~\ref{eq:loss_SE_function}.
\vspace{-2mm}
\begin{equation}
\text{Loss}_{\text{SE}} = \text{WSDR} + \text{CMS} + \text{MRS}
\label{eq:loss_SE_function}
\vspace{-2mm}
\end{equation}

The WSDR loss measures the time-domain difference between the enhanced and clean waveforms \cite{choi2018phase}. %, compensating for time delays in noisy waveforms\cite{choi2018phase}
The CMS loss measures the L1 distance between the log1p-compressed STFT magnitudes of the clean and enhanced signals, promoting accurate spectral reconstruction. The MRS loss computes STFTs of the clean and enhanced waveforms using FFT sizes of 256, 512, and 1024, and averages the L1 distance of the complex spectra across sizes to align both amplitude and phase. The SE model is evaluated under three settings: (1) \textbf{Noisy}: XRMB data with babble and non-babble noise, (2) \textbf{SE-Base}: SE model trained on noisy XRMB data using the TST method, and (3) \textbf{SISE-M}: the proposed multi-task framework jointly training SE and SI models using the TST method with the total loss defined as the sum of SE and SI losses.
The SE model is evaluated using standard objective metrics, including PESQ, CSIG, CBAK, COVL, and STOI.
%The SE model is evaluated using widely adopted objective metrics. These include: PESQ (Perceptual Evaluation of Speech Quality, range: $–$0.5 to 4.5), CSIG (Mean Opinion Score (MOS) prediction of signal distortion, range: 1 to 5), CBAK (MOS prediction of background noise intrusiveness, range: 1 to 5), COVL (MOS prediction of overall quality, range: 1 to 5), and STOI (Short-Time Objective Intelligibility, range: 0 to 1). Higher values of PESQ, CSIG, CBAK, and STOI correspond to enhanced speech that more closely resembles the original clean speech.
\vspace{-3mm}
\subsection{SI model}
\vspace{-2mm}
For the SI system, we adopt the SI architecture proposed in \cite{tabatabaee25b_interspeech}. As shown in Figure 2, the SI model estimates oral TVs in one task and the VEL TV along with three SFs in the other task. In the SI framework, each task loss is defined as a weighted combination of Pearson Correlation (PC) and Root Mean Square Error (RMSE), as shown in equation ~\ref{eq:loss_function}. The SI model’s total loss is computed as the sum of the two task losses.
\vspace{-2mm}
\begin{equation}
\text{Loss}_{\text{SI}}   = (1 - \text{PC}) + (\alpha) \cdot \text{RMSE} 
\label{eq:loss_function}
\vspace{-2mm}
\end{equation}

In equation ~\ref{eq:loss_function}, \(\alpha\) is set to 0.2. This value is determined empirically by evaluating multiple candidates and selecting the one that achieved the best performance.
We evaluate the performance of the SI model under noisy conditions across three scenarios: SI-O, SISE-P, and SISE-M. The SISE-M setting was described previously in Section 3.1. The remaining scenarios are defined as follows: (1) \textbf{SI-O}: The SI model from \cite{tabatabaee25b_interspeech}, trained on clean data without adaptation to noisy conditions, is used as the baseline. (2) \textbf{SISE-P}: The SE-Base model is used as a pre-processing step for the SI model, performing speech enhancement on input audio. Subsequently, the SI model is trained using the TST strategy.
For SI model evaluation, the Pearson Product-Moment Correlation (PPMC) is computed between the SI estimates and the ground-truth.
\vspace{-3mm}
\section{Results and Discussion}

\begin{comment}
\subsection{Evaluation of SI systems using different SSL pretrained models}
\begin{table*}[htbp]
\caption{PPMC score for estimated parameters of SI models on XRMB test set.}
%\scriptsize
\centering % Centers the table within the column span
\label{speaker_verification_models}

    \begin{tabular}{|c|c|c|c|c|c|c|c|c|c|c|c|c|c|c|}
        \hline
        SI System & SSL Model & LA & LP & TBCL & TBCD & TTCL & TTCD  &VP & Per & Aper & F0 & AVG. 
 Oral TVs   \\
         \hline    
        [?]  & WavLM-Large &0.9104 &0.7594&0.7981&0.8626&0.8360&0.9478& 0.9562 &0.9403 & 0.8815 &0.7470&0.8524\\
        \hline   
          SI  & HuBERT-Large & & &&&& &&&&&\\
        \hline
    SI  & Data2vec-Large  & & &&&& &&&&&\\
     \hline     
    SI & Wav2vec 2.0-Large   & & &&&& &&&&&\\
     \hline
    SI  & WavLM-Large  & & &&&& &&&&&\\
     \hline
    \end{tabular}
    \vspace{-3mm} 
\end{table*}
\end{comment}
\vspace{-2mm}
\subsection{Speech inversion}
\vspace{-2mm}
Table 2 shows the performance of SI-O, SISE-P, and SISE-M models on clean and noisy speech across four SNR levels with babble and non-babble noise. On the clean XRMB test set, SI-O outperforms SISE-P and SISE-M in average of PPMC scores. SI-O performance drops at low SNRs ($-$5 dB and 0 dB) since it was trained only on clean speech. Using SE pre-processing (SISE-P) or a multi-task framework (SISE-M) with noise-augmented XRMB data significantly improves performance across all SNRs and noise types compared to SI-O. Specifically, at $-$5 dB, the SISE-M model achieves relative improvements in average PPMC scores across all 10 parameters of 80.95\% for babble noise and 38.98\% for non-babble background noise, compared to the SI-O model. 

According to the results in Table 2, SISE-M achieves the highest average PPMC across all 10 parameters, except at 10 dB (both noise types) and 5 dB and 0 dB for babble noise, where it matches SISE-P. SISE-M achieves the highest PPMC scores in noisy conditions while maintaining strong performance on clean speech, showing that multi-task learning with SE and a shared SSL backbone yields robust results in both scenarios.
\vspace{-3mm}
\subsection{Speech enhancement}
\vspace{-2mm}
As shown in Table 3, incorporating the SI task within a multi-task framework improves SE performance across all SNRs and noise types, compared to fine-tuning the SE model on noisy data without leveraging SI (SE-Base). Although our primary aim is to enhance SI performance, it is encouraging that SISE-M outperforms SE-Base, demonstrating that multi-task learning improves the performance of both SI and SE systems.
\begin{figure}[htbp]
    \hfill
    \vspace{-3mm}
    \includegraphics[width=0.49\textwidth, height=0.25\textheight]{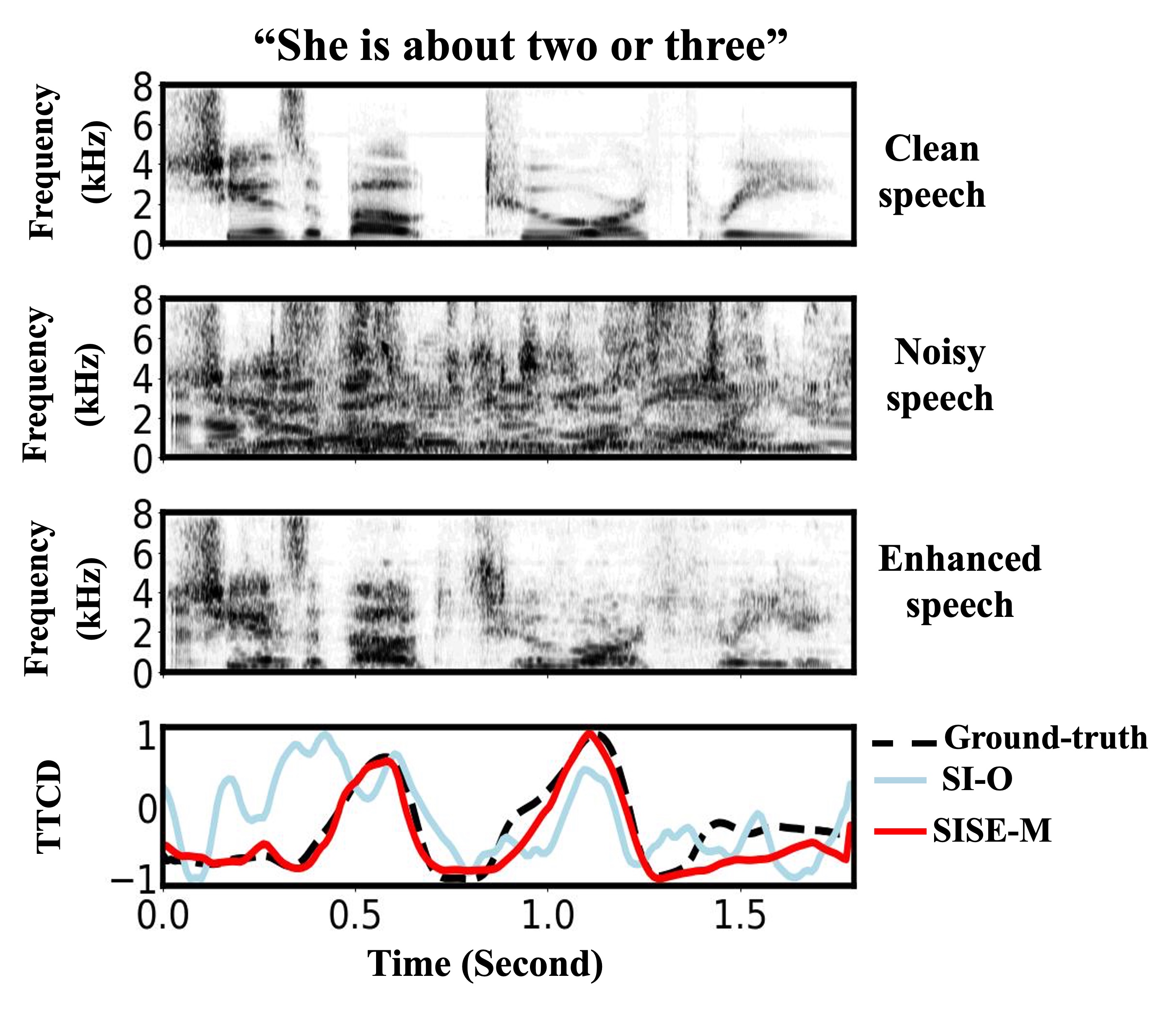}  % Adjust width as needed
     \vspace{-8mm}
    \caption{Spectrograms of clean, noisy, and enhanced speech obtained from SISE-M model. The noisy speech corresponds to an XRMB test sample corrupted by babble noise at -5 dB SNR. TTCD estimates from the SI-O and SISE-M models for the noisy speech are shown alongside the ground-truth.}    
    \label{fig:si}
    \vspace{-2mm} 
\end{figure}

Figure 3 presents spectrograms of clean, noisy, and enhanced speech processed by the SISE-M model. The noisy speech is taken from a randomly selected XRMB test sample corrupted with babble noise at $-$5 dB SNR. In addition to the spectrograms, the figure compares TTCD estimates from the SI-O baseline and the proposed SISE-M model with the ground-truth measurements. Tongue raising is observed during the “she is”, the /t/ at the end of “about” and beginning of “two”, and /r/ in “or” and during “three”. These results demonstrate that the SISE-M model accurately captures the articulatory constriction patterns, closely matching the ground-truth under challenging noisy conditions.
% SISE-M improves the PESQ, CSIG, CBAK, COVL, and STOI scores for this example from 1.02, 1.17, 1.39, 1.0, and 0.49 to 1.18, 2.77, 2.08, 1.95, and 0.73, respectively.
\vspace{-2mm}
\begin{table}[htbp]
\caption{PPMC scores for the estimated parameters of the SI-O, SISE-P, and SISE-M models. The best results are shown in bold. Per: periodicity; Aper: aperiodicity; Avg all: average PPMC score computed across all 10 parameters.}
\footnotesize
\setlength{\tabcolsep}{0.8pt}
\renewcommand{\arraystretch}{1.3}
\centering
\begin{tabular}{l|l|cccccccccc|c}
\hline
\makecell{SNR\\(dB)} & Model & LA & LP & TBCL & TBCD & TTCL & TTCD & VEL & Per & Aper & F0 & \makecell{Avg\\all}  \\
\hline
\multicolumn{12}{c}{\hspace{2cm} clean XRMB test set} \\ \hline
\multirow{3}{*}{-} & SI-O&0.91 &0.76&0.80&0.86&0.84& 0.95 & 0.96 & 0.94 & 0.88 & 0.75 &\textbf{0.87}\\ %0.85
 % & SI-F  & 0.91&0.75&0.75&0.84&0.78& 0.94 &0.97&0.95&0.89&0.78&0.86\\ %0.83
  & SISE-P &0.91 &0.76&0.71&0.84&0.79&  0.94&0.97&0.95&0.89&0.76&0.85\\ %0.83
 & SISE-M &0.91 &0.74&0.76&0.85&0.79& 0.94 &0.97&0.95&0.89&0.77&0.86\\ \hline %0.83
\multicolumn{12}{c}{\hspace{2cm}  XRMB with non-babble noise test set } \\ \hline
\multirow{3}{*}{-5} & SI-O  &0.65 &0.50&0.55&0.59&0.59&0.71  &0.57&0.66&0.57&0.52&0.59\\ % 0.60
 %  & SI-F  & 0.89&0.72&0.71&0.80&0.74&0.91  &0.92&0.91&0.83&0.73&0.82\\%0.80
   & SISE-P  & 0.89&0.72&0.67&0.80&0.75&  0.91&0.92&0.91&0.83&0.71&0.81\\%0.79
 & SISE-M  &0.89 &0.71&0.72&0.81&0.77&  0.91&0.92&0.90&0.83&0.73&\textbf{0.82}\\ \hline %0.80
\multirow{3}{*}{0} & SI-O & 0.81&0.64&0.69&0.73&0.74& 0.86 &0.76&0.82&0.74&0.63&0.74\\%0.74
  % & SI-F  &0.90 &0.74&0.72&0.82&0.76& 0.93 &0.95&0.93&0.86&0.75&0.84\\%0.81
   & SISE-P&0.90 &0.74&0.69&0.82&0.77& 0.93 &0.95&0.94&0.86&0.73&0.83\\ %0.81
 & SISE-M & 0.90&0.73&0.74&0.83&0.78& 0.93 &0.95&0.93&0.86&0.75&\textbf{0.84}\\ \hline %0.82
\multirow{3}{*}{5} & SI-O & 0.87&0.70&0.74&0.80&0.80& 0.91 &0.85&0.89&0.81&0.68&0.81\\%0.80
 %  & SI-F  & 0.91&0.74&0.73&0.83&0.77& 0.93 &0.96&0.94&0.88&0.76&0.85\\ %0.82
   & SISE-P &0.91 &0.75&0.70&0.83&0.78&0.93  &0.96&0.95&0.87&0.74&0.84\\ %0.82
 & SISE-M  &0.90 &0.73&0.74&0.84&0.78&  0.93&0.96&0.94&0.87&0.76&\textbf{0.85}\\ \hline %0.82
\multirow{3}{*}{10} & SI-O&0.90 &0.73&0.76&0.83&0.83& 0.93 &0.88&0.92&0.83&0.71&0.83\\%0.83
 % & SI-F   & 0.91&0.75&0.73&0.83&0.77& 0.93 &0.97&0.95&0.88&0.77&0.85\\%0.82
   & SISE-P &0.91 &0.75&0.71&0.83&0.78&0.93  &0.97&0.95&0.88&0.75&\textbf{0.85}\\ %0.82
 & SISE-M &0.91 &0.74&0.75&0.84&0.78&0.94  &0.96&0.94&0.88&0.76&\textbf{0.85}\\ \hline
\multicolumn{12}{c}{\hspace{2cm} XRMB with babble noise test set  } \\ \hline
\multirow{3}{*}{-5} & SI-O &0.48 &0.36&0.40&0.45&0.41&0.52 &0.39&0.42&0.34&0.39&0.42\\%0.44
%   & SI-F  & 0.81&0.67&0.65&0.74&0.70& 0.85 &0.82&0.81&0.74&0.65&0.74\\%0.74
   & SISE-P & 0.81&0.66&0.62&0.73&0.69&0.84  &0.80&0.80&0.72&0.62&0.73\\ % 0.73
 & SISE-M  &0.83 &0.68&0.65&0.75&0.71&0.86  &0.84&0.82&0.75&0.66&\textbf{0.76}\\ \hline%0.75
\multirow{3}{*}{0} & SI-O  & 0.72&0.55&0.59&0.64&0.62& 0.75 &0.61&0.68&0.60&0.53&0.63\\%0.65

 %  & SI-F  &0.90 &0.74&0.72&0.82&0.76& 0.93 &0.94&0.92&0.85&0.74&0.83\\%0.81
   & SISE-P &0.90 &0.74&0.70&0.82&0.77&  0.92&0.94&0.92&0.85&0.72&\textbf{0.83}\\ %0.81
 & SISE-M  &0.90 &0.74&0.72&0.82&0.76&  0.93&0.94&0.92&0.85&0.74&\textbf{0.83}\\%0.81
 \hline
\multirow{3}{*}{5} & SI-O & 0.85&0.68&0.72&0.78&0.77& 0.89 &0.79&0.85&0.77&0.64&0.77\\%0.78
%  & SI-F &0.90 &0.74&0.73&0.83&0.77& 0.93 &0.96&0.94&0.87&0.76&0.84\\%0.82
   & SISE-P & 0.91&0.75&0.70&0.83&0.78& 0.93 &0.96&0.94&0.87&0.74&\textbf{0.84}\\ %0.82
 & SISE-M& 0.90&0.74&0.73&0.83&0.77&  0.93&0.96&0.94&0.87&0.76&\textbf{0.84}\\ \hline%0.82
\multirow{3}{*}{10} & SI-O  & 0.89&0.72&0.75&0.82&0.81& 0.92 &0.87&0.91&0.82&0.68&0.82\\%0.82
%  & SI-F & 0.91&0.74&0.74&0.83&0.77&0.93  &0.96&0.95&0.88&0.76&0.85\\%0.82
   & SISE-P &0.91 &0.75&0.71&0.83&0.78&  0.93&0.97&0.95&0.88&0.75&\textbf{0.85}\\%0.82
 & SISE-M &0.91 &0.74&0.75&0.84&0.79& 0.94 &0.96&0.94&0.88&0.76&\textbf{0.85}\\ \hline %0.83
 
    \end{tabular}
    \vspace{-3mm} 
\end{table}
%\vspace{-1mm}
\subsection{Performance under babble versus non-babble noise}
\vspace{-2mm}
Both the SE and SI models showed lower performance for babble background noise compared to non-babble background noise, indicating that babble noise poses a more challenging environment for both tasks, as it interferes more with the target speech signal. Specifically, under noisy conditions ($-$5 dB), the best-performing model, SISE-M, shows a 7.89\% relative improvement in the average PPMC score for the SI task with non-babble background noise compared to babble noise, and a 16.55\% relative improvement in PESQ for the SE task under the same conditions.

\vspace{-2mm} 
\begin{table}[htbp]
\vspace{-1mm} 
\caption{Comparison of SE-Base and SISE-M performance under the noisy condition at various SNR levels. The best results are shown in bold.}
\footnotesize
\setlength{\tabcolsep}{7pt}
\renewcommand{\arraystretch}{1.1}
\centering
\begin{tabular}{c|c|ccccc}
\hline
\makecell{SNR\\(dB)} & Model  &PESQ & CSIG &CBAK &COVL& STOI   \\
\hline
\multicolumn{7}{c}{\hspace{2cm}  XRMB with non-babble noise test set  } \\ \hline
\multirow{3}{*}{-5} 
&Noisy &1.10 & 1.63&1.70&1.32&0.65\\
& SE-Base&1.56 &3.21&2.37&2.38&0.83\\
 & SISE-M &\textbf{1.62} &\textbf{3.29}&\textbf{2.41}&\textbf{2.46}& \textbf{0.84}\\ \hline
\multirow{3}{*}{0} %& Baseline  &1.61 &0.68\\
&Noisy & 1.15& 2.00&1.95&1.53&0.74\\
& SE-Base&1.97 &3.69&2.76&2.85&0.89\\
 & SISE-M & \textbf{2.05}&\textbf{3.77}&\textbf{2.81}&\textbf{2.94}& \textbf{0.90}\\ \hline
\multirow{3}{*}{5} %& Baseline  &2.05 & 0.90\\
&Noisy & 1.25&2.47&2.28&1.84&0.82 \\
& SE-Base &2.45 &4.11&3.19&3.32&\textbf{0.93}\\
 & SISE-M  & \textbf{2.55}&\textbf{4.20}&\textbf{3.24}&\textbf{3.42}&\textbf{0.93}\\ \hline
\multirow{3}{*}{10} %& Baseline  & 2.61& 0.95\\
&Noisy & 1.47& 2.96&2.67&2.23&0.89\\
& SE-Base& 2.91&4.46&3.62&3.74&0.95\\
 & SISE-M  & \textbf{3.03}&\textbf{4.56}&\textbf{3.68}&\textbf{3.86}& \textbf{0.96}\\ \hline
 
\multicolumn{7}{c}{\hspace{2cm} XRMB with babble noise test set  } \\ \hline
\multirow{3}{*}{-5} 
&Noisy & 1.09&1.63&1.65&1.30&0.54\\
& SE-Base & 1.34&2.88&2.14&2.07&0.72\\
 & SISE-M &\textbf{1.39}&\textbf{2.99}&\textbf{2.18}&\textbf{2.16}& \textbf{0.74}\\ \hline
\multirow{3}{*}{0} %& Baseline  &1.56 & 0.71\\
&Noisy & 1.13&2.02&1.91&1.51&0.66\\
& SE-Base &1.81 &3.58&2.65&2.70&\textbf{0.86}\\\
 & SISE-M  &\textbf{1.86}&\textbf{3.65}&\textbf{2.67}&\textbf{2.77}&\textbf{0.86}\\ \hline
\multirow{3}{*}{5} %& Baseline  &2.04 &0.90 \\
&Noisy &1.24 &2.54&2.25&1.86& 0.77\\
& SE-Base & 2.33&4.04&3.11&3.22&\textbf{0.91}\\
 & SISE-M & \textbf{2.39}&\textbf{4.11}&\textbf{3.14}&\textbf{3.29}& \textbf{0.91}\\ \hline
\multirow{3}{*}{10} %& Baseline  & 2.53& 0.94\\
&Noisy &1.51 &3.11&2.69&2.31&0.86\\
& SE-Base &2.78 &4.39&3.54&3.63&\textbf{0.94}\\
 & SISE-M &\textbf{2.88}&\textbf{4.49}&\textbf{3.60}&\textbf{3.74}&\textbf{0.94}\\ \hline
 
    \end{tabular}
    \vspace{-1mm} 
  \end{table}

\vspace{-2mm}
\section{Conclusions}
\vspace{-2mm}
We proposed a multi-task learning approach that leverages a unified SSL speech representation framework to jointly optimize SE and SI, with the primary goal of improving SI robustness in noisy environments. Experimental results show that this joint training strategy consistently outperforms separate training, where SE is used solely as a pre-processing step for SI, across diverse noise types and SNR levels. Moreover, the multi-task framework not only strengthens SI performance but also enhances SE, highlighting its effectiveness as a practical solution for real-world noisy scenarios. Finally, our findings suggest that SSL speech representations provide a powerful foundation for simultaneously performing SE and SI tasks.
\begin{comment}
\begin{table}[h!]
\centering
\begin{tabular}{|l|c|}
\hline
\textbf{Parameter} & \textbf{PPMC Score} \\
\hline
\multicolumn{2}{|c|}{\textbf{Oral TVs}} \\
\hline
LA                & 0.9203 \\
LP                & 0.7697 \\
TBCL              & 0.7859 \\
TBCD              & 0.8505 \\
TTCL              & 0.8409 \\
TTCD              & 0.9473 \\
\textbf{Average (TVs)} & \textbf{0.8524} \\
\hline
\multicolumn{2}{|c|}{\textbf{Other Parameters}} \\
\hline
VP        & 0.9454 \\
Periodicity       & 0.9571 \\
Aperiodicity      & 0.8493 \\
Pitch             & 0.7452 \\
\hline
\end{tabular}
\label{tab:feature_scores}
\end{table}
\end{comment}

%\bibliographystyle{IEEEtran}
%\bibliography{reference}   
\makeatletter
\patchcmd{\thebibliography}{\settowidth}{%
  \setlength{\itemsep}{0pt}%
  \setlength{\parskip}{0pt}%
  \setlength{\parsep}{0pt}%
  \setlength{\topsep}{0pt}%
  \small 
  \settowidth
}{}{}
\makeatother

\bibliographystyle{IEEEtran}
\bibliography{reference}

@book{stevens2000acoustic,
  title={Acoustic phonetics},
  author={Stevens, Kenneth N},
  volume={30},
  year={2000},
  publisher={MIT press}
}

@inproceedings{botinhao2016investigating,
  title={Investigating RNN-based speech enhancement methods for noise-robust text-to-speech},
  author={Botinhao, Cassia Valentini and Wang, Xin and Takaki, Shinji and Yamagishi, Junichi},
  booktitle={9th ISCA speech synthesis workshop},
  pages={159--165},
  year={2016}
}

@article{chen2022wavlm,
  title={Wavlm: Large-scale self-supervised pre-training for full stack speech processing},
  author={Chen, Sanyuan and Wang, Chengyi and Chen, Zhengyang and Wu, Yu and Liu, Shujie and Chen, Zhuo and Li, Jinyu and Kanda, Naoyuki and Yoshioka, Takuya and Xiao, Xiong and others},
  journal={IEEE Journal of Selected Topics in Signal Processing},
  volume={16},
  number={6},
  pages={1505--1518},
  year={2022},
  publisher={IEEE}
}

@inproceedings{huang2022investigating,
  title={Investigating self-supervised learning for speech enhancement and separation},
  author={Huang, Zili and Watanabe, Shinji and Yang, Shu-wen and Garc{\'\i}a, Paola and Khudanpur, Sanjeev},
  booktitle={ICASSP 2022-2022 IEEE International Conference on Acoustics, Speech and Signal Processing (ICASSP)},
  pages={6837--6841},
  year={2022},
  organization={IEEE}
}

@inproceedings{cho2023evidence,
  title={Evidence of vocal tract articulation in self-supervised learning of speech},
  author={Cho, Cheol Jun and Wu, Peter and Mohamed, Abdelrahman and Anumanchipalli, Gopala K},
  booktitle={ICASSP 2023-2023 IEEE International Conference on Acoustics, Speech and Signal Processing (ICASSP)},
  pages={1--5},
  year={2023},
  organization={IEEE}
}

@inproceedings{khan2024exploiting,
  title={Exploiting consistency-preserving loss and perceptual contrast stretching to boost ssl-based speech enhancement},
  author={Khan, Muhammad Salman and La Quatra, Moreno and Hung, Kuo-Hsuan and Fu, Szu-Wei and Siniscalchi, Sabato Marco and Tsao, Yu},
  booktitle={2024 IEEE 26th International Workshop on Multimedia Signal Processing (MMSP)},
  pages={1--6},
  year={2024},
  organization={IEEE}
}

@inproceedings{choi2018phase,
  title={Phase-aware speech enhancement with deep complex u-net},
  author={Choi, Hyeong-Seok and Kim, Jang-Hyun and Huh, Jaesung and Kim, Adrian and Ha, Jung-Woo and Lee, Kyogu},
  booktitle={International Conference on Learning Representations},
  year={2018}
}

@inproceedings{attia2024improving,
  title={Improving speech inversion through self-supervised embeddings and enhanced tract variables},
  author={Attia, Ahmed Adel and Siriwardena, Yashish M and Espy-Wilson, Carol},
  booktitle={2024 32nd European Signal Processing Conference (EUSIPCO)},
  pages={306--310},
  year={2024},
  organization={IEEE}
}

@article{siriwardena2024speaker,
  title={Speaker-independent speech inversion for recovery of velopharyngeal port constriction degree},
  author={Siriwardena, Yashish M and Boyce, Suzanne E and Tiede, Mark K and Oren, Liran and Fletcher, Brittany and Stern, Michael and Espy-Wilson, Carol Y},
  journal={The Journal of the Acoustical Society of America},
  volume={156},
  number={2},
  pages={1380--1390},
  year={2024},
  publisher={AIP Publishing}
}

@inproceedings{reddy2021interspeech,
  title={INTERSPEECH 2021 Deep Noise Suppression Challenge},
  author={Reddy, Chandan KA and Dubey, Harishchandra and Koishida, Kazuhito and Nair, Arun and Gopal, Vishak and Cutler, Ross and Braun, Sebastian and Gamper, Hannes and Aichner, Robert and Srinivasan, Sriram},
  booktitle={INTERSPEECH},
  year={2021}
}

@inproceedings{panayotov2015librispeech,
  title={Librispeech: an asr corpus based on public domain audio books},
  author={Panayotov, Vassil and Chen, Guoguo and Povey, Daniel and Khudanpur, Sanjeev},
  booktitle={2015 IEEE international conference on acoustics, speech and signal processing (ICASSP)},
  pages={5206--5210},
  year={2015},
  organization={IEEE}
}

@article{nagrani2020voxceleb,
  title={Voxceleb: Large-scale speaker verification in the wild},
  author={Nagrani, Arsha and Chung, Joon Son and Xie, Weidi and Zisserman, Andrew},
  journal={Computer Speech \& Language},
  volume={60},
  pages={101027},
  year={2020},
  publisher={Elsevier}
}

@inproceedings{siriwardena2023audio,
  title={Audio data augmentation for acoustic-to-articulatory speech inversion},
  author={Siriwardena, Yashish M and Attia, Ahmed Adel and Sivaraman, Ganesh and Espy-Wilson, Carol},
  booktitle={2023 31st European Signal Processing Conference (EUSIPCO)},
  pages={301--305},
  year={2023},
  organization={IEEE}
}

@article{shahrebabaki2021acoustic,
  title={Acoustic-to-articulatory mapping with joint optimization of deep speech enhancement and articulatory inversion models},
  author={Shahrebabaki, Abdolreza Sabzi and Salvi, Giampiero and Svendsen, Torbj{\o}rn and Siniscalchi, Sabato Marco},
  journal={IEEE/ACM Transactions on Audio, Speech, and Language Processing},
  volume={30},
  pages={135--147},
  year={2021},
  publisher={IEEE}
}

@inproceedings{tabatabaee25b_interspeech,
  title     = {{Enhancing Acoustic-to-Articulatory Speech Inversion by Incorporating Nasality}},
  author    = {Saba Tabatabaee and Suzanne Boyce and Liran Oren and Mark Tiede and Carol Espy-Wilson},
  year      = {2025},
  booktitle = {{Interspeech 2025}},
  pages     = {325--329},
  doi       = {10.21437/Interspeech.2025-2387},
  issn      = {2958-1796},
}

@article{tabatabaee2025acoustic,
  title={Acoustic to Articulatory Speech Inversion for Children with Velopharyngeal Insufficiency},
  author = {{Tabatabaee}, Saba and {Boyce}, Suzanne and {Oren}, Liran and {Tiede}, Mark and {Espy-Wilson}, Carol},
  journal={arXiv preprint arXiv:2509.09489},
  year={2025}
}

@article{shi2024waveform,
  title={Waveform-domain speech enhancement using spectrogram encoding for robust speech recognition},
  author={Shi, Hao and Mimura, Masato and Kawahara, Tatsuya},
  journal={IEEE/ACM Transactions on Audio, Speech, and Language Processing},
  volume={32},
  pages={3049--3060},
  year={2024},
  publisher={IEEE}
}

@inproceedings{tzeng2025noise,
  title={Noise-Robust Speech Emotion Recognition Using Shared Self-Supervised Representations with Integrated Speech Enhancement},
  author={Tzeng, Jing-Tong and Leem, Seong-Gyun and Salman, Ali N and Lee, Chi-Chun and Busso, Carlos},
  booktitle={ICASSP 2025-2025 IEEE International Conference on Acoustics, Speech and Signal Processing (ICASSP)},
  pages={1--5},
  year={2025},
  organization={IEEE}
}

@article{benway2025perceptual,
  title={Perceptual Ratings Predict Speech Inversion Articulatory Kinematics in Childhood Speech Sound Disorders},
  author={Benway, Nina R and Tabatabaee, Saba and Wang, Dongliang and Munson, Benjamin and Preston, Jonathan L and Espy-Wilson, Carol},
  journal={arXiv preprint arXiv:2507.01888},
  year={2025}
}

@inproceedings{benway25_interspeech,
  title     = {{Subtyping Speech Errors in Childhood Speech Sound Disorders with Acoustic-to-Articulatory Speech Inversion}},
  author    = {Nina R Benway and Saba Tabatabaee and Benjamin Munson and Jonathan Preston and Carol Espy-Wilson},
  year      = {2025},
  booktitle = {{Interspeech 2025}},
  pages     = {2800--2804},
  doi       = {10.21437/Interspeech.2025-339},
  issn      = {2958-1796},
}

@article{Snyder2015MUSANAM,
  title={MUSAN: A Music, Speech, and Noise Corpus},
  author={David Snyder and Guoguo Chen and Daniel Povey},
  journal={ArXiv},
  year={2015},
  volume={abs/1510.08484},
  url={https://api.semanticscholar.org/CorpusID:15676318}
}

@article{deshmukh2005use,
  title={Use of temporal information: Detection of periodicity, aperiodicity, and pitch in speech},
  author={Deshmukh, Om and Espy-Wilson, Carol Y and Salomon, Ariel and Singh, Jawahar},
  journal={IEEE Transactions on Speech and Audio Processing},
  volume={13},
  number={5},
  pages={776--786},
  year={2005},
  publisher={IEEE}
}

@article{westbury1994speech,
  title={Speech Production Database User’S Handbook},
  author={Westbury, John R},
  journal={IEEE Personal Communications-IEEE Pers. Commun., vol. 0, no},
  year={1994}
}

@article{browman1986towards,
  title={Towards an articulatory phonology},
  author={Browman, Catherine P and Goldstein, Louis M},
  journal={Phonology},
  volume={3},
  pages={219--252},
  year={1986},
  publisher={Cambridge University Press}
}
\end{document}